\documentclass[reprint,aps,prl,superscriptaddress]{revtex4-2}

\usepackage{bm, physics, amsmath, amssymb, graphicx, xcolor, embedfile, comment}
\usepackage[utf8]{inputenc}
\usepackage[normalem]{ulem}
\usepackage[T1]{fontenc}
\usepackage{tikz}

\graphicspath{{figures/}}
\DeclareUnicodeCharacter{03B3}{\ensuremath{\gamma}}

\makeindex

\begin{document}

\title{Spontaneous flexoelectricity in cubic lead-halide perovskite MAPbBr$_3$}

\author{Dmytro Rak}
\affiliation{IST Austria (Institute of Science and Technology Austria), Am Campus 1, 3400 Klosterneuburg, Austria}
\affiliation{Institute of Experimental Physics, Slovak Academy of Sciences, Watsonova 47, 040 01 Košice, Slovakia}
     
\author{Dusan Lorenc}
\affiliation{IST Austria (Institute of Science and Technology Austria), Am Campus 1, 3400 Klosterneuburg, Austria}
    
\author{Ayan A. Zhumekenov}
\affiliation{King Abdullah University of Science and Technology, Thuwal 23955, Saudi Arabia} 
    
\author{Osman M. Bakr}
\affiliation{Center for Renewable Energy and Storage Technologies (CREST), Division of Physical Science and Engineering (PSE), King Abdullah University of Science and Technology, Thuwal 23955-6900, Kingdom of Saudi Arabia}

\author{Zhanybek Alpichshev}
\email{alpishev@ist.ac.at}
\affiliation{IST Austria (Institute of Science and Technology Austria), Am Campus 1, 3400 Klosterneuburg, Austria}

\begin{abstract}
Lead-halide perovskites exhibit remarkable efficiency in photovoltaics, driven by exceptionally long carrier diffusion lengths and recombination times. Paradoxically, this performance persists even in defect-rich, solution-grown samples. Here, we use a suite of optical and charge transport measurements to reveal that key optoelectronic properties of perovskites arise from localized flexoelectric polarization confined to the interfaces between domains of spontaneous strain, present even in nominally cubic single crystals. This insight provides a microscopic link between structural composition and charge transport in these materials, reconciling conflicting prior observations and offering new design principles for perovskite-based solar cells.
\end{abstract}

\maketitle

\section{Introduction}

Halide perovskites have come forward to the limelight as prospective future-generation photovoltaic materials, offering significant advantages over conventional technologies, such as low manufacturing costs and solution processibility. The efficiency of the perovskite solar cells has improved tremendously over the last decade and has now reached 26.7\% \cite{NREL2023,Min2021} for single-junction cells. Lead-halide perovskites (LHPs) are particularly interesting for solar energy harvesting applications. The remarkable photovoltaic performance of these materials is a consequence of a combination of factors, such as high optical absorption coefficients \cite{DeWolf2014, Yang2015}, high defect tolerance \cite{Kim2014, Dong2015, Saidaminov2015}, and long charge carrier lifetimes \cite{Dong2015, Wehrenfennig2014} and diffusion lengths \cite{Dong2015, Stranks2013}. However, despite intensive research, the nature of the optoelectronic properties underlying such astounding performance is unclear.

LHPs such as MAPbI$_3$ and MAPbBr$_3$ (MA = CH$_3$NH$^+_3$) were thoroughly studied to understand the nature of their superior optoelectronic properties. It was realized early on that the diffusion coefficients in LHPs are not remarkable at all, and that the exceptional diffusion lengths -- critical for photovoltaic applications -- are mainly due to exceptionally long photocarrier recombination times (\textit{e.g.} \cite{Dong2015}). Several explanations were proposed, such as strong spin-orbit coupling leading to the Rashba splitting \cite{Kepenekian2015, Niesner2016, Lafalce2022}, a bulk photovoltaic effect resulting from ferroelectric ordering \cite{Frost2014, Rakita2017, Garten2019, Gao2019}, or reduced scattering of the charge carriers protected by the formation of large polarons \cite{Frost2017, Miyata2018}. Particular attention was paid to the possibility of ferroelectric ordering, which would explain the long lifetimes and diffusion lengths of the charge carriers. The idea of ferroelectric ordering is generally tempting, considering the oxide perovskites are well-known ferroelectrics. However, most of the structural studies of MAPbI$_3$ and MAPbBr$_3$ do not support this hypothesis. At room temperature, their space groups were identified \cite{Poglitsch1987, Mashiyama2003, Baikie2015} as non-polar tetragonal $I4/mcm$ and cubic $Pm\bar{3}m$, respectively, with ferroelectricity forbidden by symmetry. At the same time, some works identified the structure of the tetragonal MAPbI$_3$ as $I4cm$ polar group \cite{Stoumpos2013, Dang2015}. With recent works putting forward compelling evidence of the ferroelectric ordering in the tetragonal phase of MAPbI$_3$ \cite{Rakita2017, Garten2019} and MAPbBr$_3$ \cite{Gao2019}, the puzzling nature of the tetragonal phase has become a subject of extensive debate. Meanwhile, the cubic phase has received much less attention even though the tetragonal-to-cubic phase transition has little effect on the performance of MAPbI$_3$ in solar applications \cite{Zhang2015}, indicating that the mechanism underlying its unique properties remains unaffected.

In this paper, we investigate the optoelectronic properties of the room-temperature polymorph of MAPbBr$_3$, previously identified as cubic. Our key finding is the discovery of local flexoelectric polarization confined to the boundaries between microscopic domains of unequal strain, with resultant local electric fields generating long-lived photocurrents under zero bias. We find that the spatial separation of photocarriers at the flexoelectrically-polarized domain walls and the consequent exponential suppression of their recombination naturally explains many of the previously reported unexpected aspects of charge dynamics in LHPs. We confirm the complex domain structure of the nominally cubic phase by visualizing the domain walls using an original electrochemical staining technique. Our results provide new insights into the structural complexity of lead-halide perovskites and open new avenues for designing and optimizing their application in photovoltaic technology.

\begin{figure*}
\includegraphics[width=\textwidth]{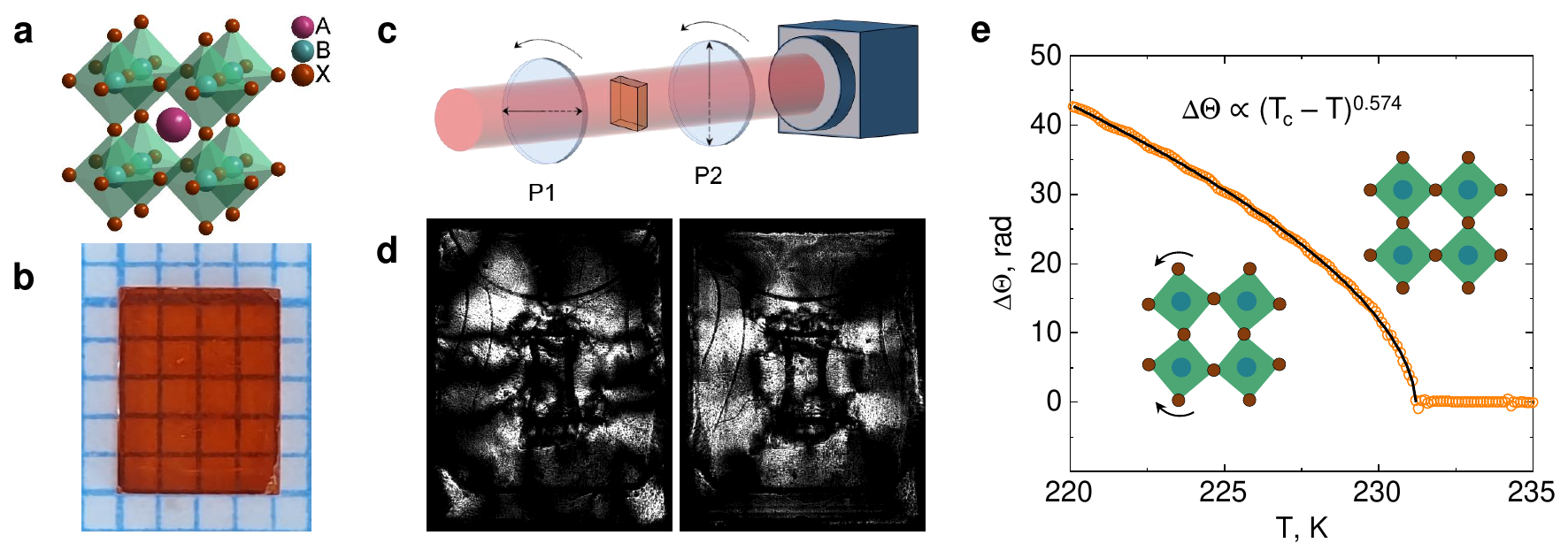}
\caption{
\textbf{Birefringence in the nominally cubic high-temperature phase of MAPbBr$_3$.}
 \textbf{a}, Crystal structure of cubic perovskite with the general chemical formula ABX$_3$.
\textbf{b}, A typical MAPbBr$_3$ crystal grown from solution using the inverse temperature crystallization technique.
\textbf{c}, Schematic of the crossed-polarizer setup used to visualize the birefringence in the high-temperature phase of MAPbBr$_3$.
\textbf{d}, Transmission images of the MAPbBr$_3$ crystal shown in \textbf{b} placed in the crossed-polarizer setup. The incident He-Ne light is polarized at 0$^\circ$ (left) and 45$^\circ$ (right) relative to the [010] crystal axis.
\textbf{e}, Cumulative phase retardation $\Delta \Theta$ as a function of temperature across the cubic-to-tetragonal phase transition in MAPbBr$_3$ single crystal. The solid black line represents the best fit to the data above the nominal critical temperature $T_c = 231.2^\circ$C of the form $\Delta \Theta \propto |T_c-T|^{\alpha}$. Inset images show the tilting pattern of PbBr$_6$ octahedra associated with the phase transition.
}
\label{fig:fig1}
\end{figure*}

\section{Results}

\subsection{Birefringence in cubic MAPbBr$_3$}

Our investigation begins with the observation of an unexpected optical anisotropy in the room-temperature phase of MAPbBr$_3$. Similar to other LHPs, the basic structure of MAPbBr$_3$ is composed of corner-sharing BX$_6$ (Fig.\ref{fig:fig1}a) octahedra, forming transparent rectangular-shaped crystals with a nominally cubic symmetry above $T \approx 230K$ \cite{Bari2021}. However, a careful study of generic crystals demonstrates this is not the case in general. Fig.\ref{fig:fig1}b shows a typical MAPbBr$_3$ crystal grown from solution by inverse temperature crystallization \cite{Saidaminov2015}. To examine its optical properties, the crystal was placed in the rotating crossed-polarizer setup (Fig.\ref{fig:fig1}c) and illuminated by a 632.8$n$m He-Ne laser beam normal to the (100) plane while changing the angle of polarization relative to the [010] crystal axis. The rotation of the crossed polarizer was captured on video \cite{supp}. Fig.\ref{fig:fig1}d shows transmitted light images of MAPbBr$_3$ crystal illuminated with light polarized at 0$^\circ$ and 45$^\circ$, respectively, revealing the complex birefringent structure. This is an unexpected observation since MAPbBr$_3$ is supposed to be in cubic phase at room temperature \cite{OnodaYamamuro1990}, which is incompatible with the observed birefringence. This definitively indicates that the crystal symmetry of the MAPbBr$_3$ sample is lower than the nominal $Pm\bar{3}m$ space group. While the birefringence of nominally cubic MAPbBr$_3$ is commonly observed, this incongruity is not systematically addressed in the literature. It is typically attributed to the inferior quality of the solution-processed crystals \cite{Song2023}, but the exact nature of this symmetry lowering, however, is yet to be understood.

The cubic-to-tetragonal transition in MAPbBr$_3$ is identified as a close-to-second-order first-order displacement-type phase transition driven by the rotation of BX$_6$ octahedra \cite{OnodaYamamuro1990, Mashiyama2003} (see inset to Fig.\ref{fig:fig1}e). Reflecting the structural change, the electric susceptibility tensor is also modified at the phase transition which is manifested in the corresponding change in the optical refractive index ellipsoid. Birefringence, therefore, provides information on the onset of phase transition and can be used as an indicator of the order parameter of the phase transition. X-ray diffraction studies revealed that the order parameter of the cubic-to-tetragonal phase transition in MAPbBr$_3$ is the rotation angle of the PbBr$_6$ octahedra \cite{Mashiyama2003}. The idea that the same order parameter drives the formation of the distorted cubic phase at room temperature seemed plausible, which is why we proceeded with measuring the temperature dependence of birefringence across the phase transition. The critical behavior was evaluated by measuring the temperature dependency of the Stokes parameters of a polarized near-infrared beam passing through the MAPbBr$_3$ crystal placed inside the cryostat \cite{supp}, and fitting the cumulative phase retardation acquired by the beam to a power law $\Delta \Theta \propto |T-T_c|^{\alpha}$ (Fig.\ref{fig:fig1}e). As can be seen in the figure, the temperature dependence of birefringence retains the sharp anomaly at $T_c$, with ``high-temperature'' birefringence providing a simple offset to the critical birefringence associated with the phase transition. This indicates that the symmetry-

\begin{figure*}
\includegraphics[width=0.8\textwidth]{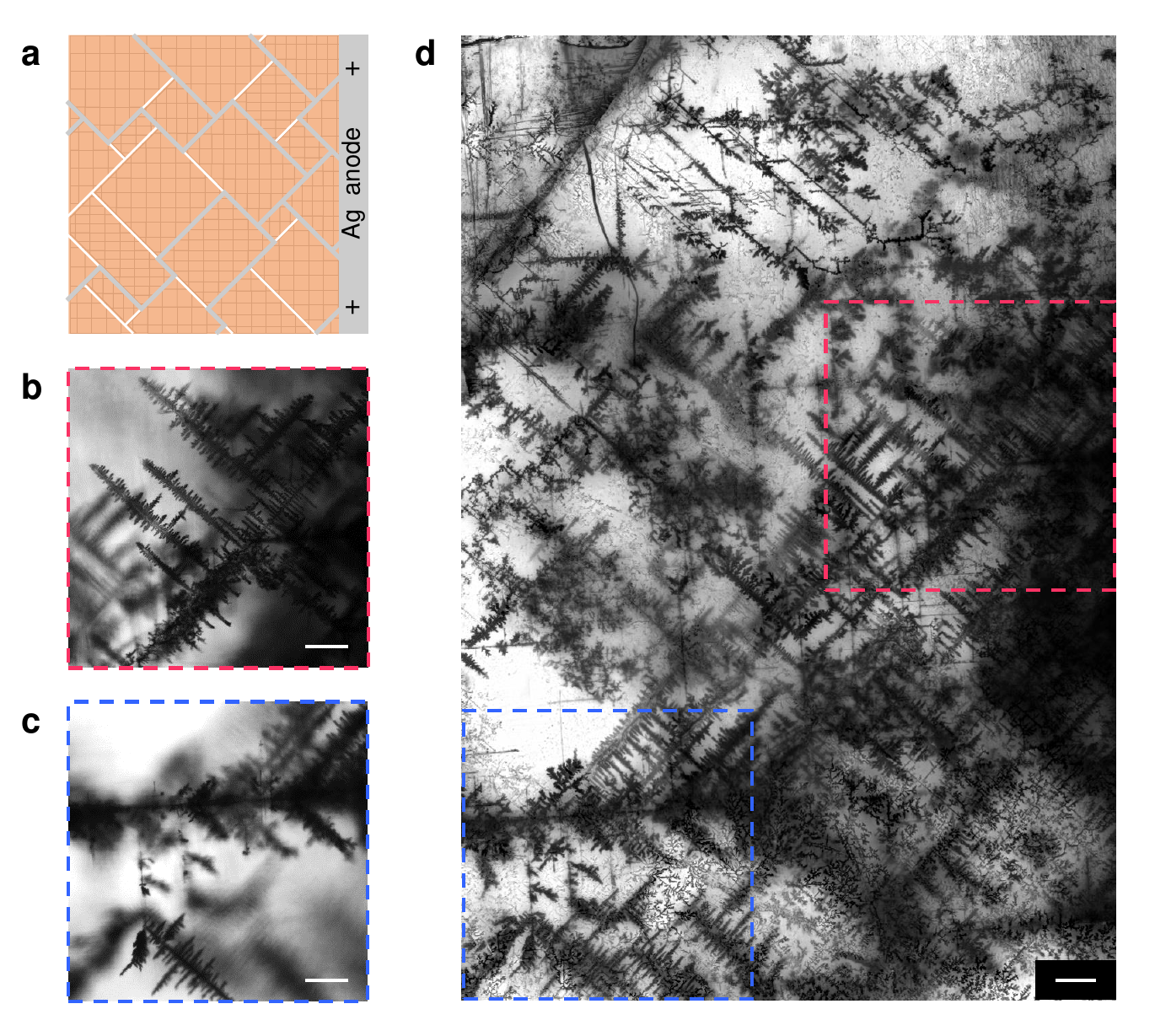}
\caption{
\textbf{Electrochemical visualization of the domain structure in MAPbBr$_3$.}
 \textbf{a}, Schematic of the experiment. The silver anode attached to the side of the sample serves as a source of silver ions. When an electric field is applied, the ions are electrophoretically injected into the crystal. The domain structure is revealed through the electrochemical reduction of silver ions to metallic silver (grey), making the domain walls (white) visible. The hatched areas represent domains with different orientations of strain and magnitudes of strain.
\textbf{b,c}, Bright-field images of silver dendrites formed in MAPbBr$_3$ monocrystal after an electric field of 24V/mm was applied for 4 hours. The images show single optical slices taken at 270$\mu$m (\textbf{b}) and 365$\mu$m (\textbf{c}) beneath the sample surface. The scale bars are 100$\mu$m.
\textbf{d}, Composite image with an extended depth of field, created by focus stacking 167 bright-field images. The dashed lines mark the positions where the corresponding single optical slices displayed in \textbf{b} and \textbf{c} were taken. The scale bar is 100$\mu$m.
}
\label{fig:fig2}
\end{figure*}

\noindent lowering of MAPbBr$_3$ at room temperature has a nature distinct from the low-temperature rotation of PbBr$_6$ octahedra in the tetragonal phase.

\subsection{Microscopic structure of local anistropy in MAPbBr$_3$}

A birefringent sample can appear dark when placed between two orthogonally crossed polarizers only if the polarization axis of one of the polarizers matches with the optical axis of the sample. In contrast, an optically active sample will always appear bright, while isotropic sample will be always dark. Our optical measurements showed that no part of the sample remains permanently bright or dark for all orientations of the mutually crossed polarizers relative to the crystal axes \cite{supp}. The data indicates that the sample is birefringent everywhere with no optical activity and with locally-defined optical axes seemingly uncorrelated with crystal directions. 

Birefringence in a nominally cubic system indicates the presence of bulk strain. In a free-standing crystal inhomogeneous strain implies the presence of defects such as dislocations or -- in case the strain is spontaneous (ferroelasticity) -- domain boundaries \cite{Landau1986}. To investigate the nature of nonuniform strain in MAPbBr$_3$, we electrophoretically inject silver ions into the sample, which in the course of diffusion preferentially cluster near defect sites. Thus, when ions are ultimately reduced to metallic silver, the defects become visible under microscope (Fig.\ref{fig:fig2}a; Methods). Fig.\ref{fig:fig2}b,c shows the silver-stained defect patterns in a typical solution-grown MAPbBr$_3$ monocrystal taken near a silver electrode attached to (010) face of the crystal \cite{supp}. Dendritic structures clearly indicate the presence of a domain wall pattern oriented at 45$^\circ$ (b) and 90$^\circ$ (c) relative to crystal axes. Domains as small as $2 \times 2\mu$m$^2$ were identified \cite{supp}. Fig.\ref{fig:fig2}d shows an image of the silver-stained sample generated by the focal plane merging of the bright-field Z-stack, revealing a complex domain structure of the distorted cubic phase. The video showing a complete scan along the Z-axis can be found in the supplementary \cite{supp}.

The microscopic images were taken shortly after the applied voltage was removed. Left unperturbed for several hours, silver structures start to dissolve, smearing the spatial pattern. Heating the crystal increases the dissolution rate and leads to the complete disappearance of silver structures over time. Interestingly, MAPbBr$_3$ electrophoretically doped with silver also demonstrates reversible photochromic effect \cite{supp}; assuming that silver bromide is formed during electrochemical staining, the mechanism of this effect is likely similar to the one occurring in silver halide-containing photochromic glasses \cite{Armistead1964}.

Electrochemical staining indicates that non-cubicity in room-temperature MAPbBr$_3$ comes in the form of microscopic ferroelastic domains each hosting different strain uniform across the domain. Importantly, in this picture strain gradients are confined to structural defects - domain walls - thus alleviating the necessity of associated stress gradients. The few-micron domain size in MAPbBr$_3$ explains the illusion of smooth gradients of birefringence in the sample shown in Fig.\ref{fig:fig1}d. The interpretation of room-temperature phase in terms of ferroelasticity is reinforced by X-ray diffraction, which confirms the previously reported $Pm\bar{3}m$ space group and demonstrates the structural homogeneity of the sample \cite{supp}.

\begin{figure*}
\includegraphics[width=\textwidth]{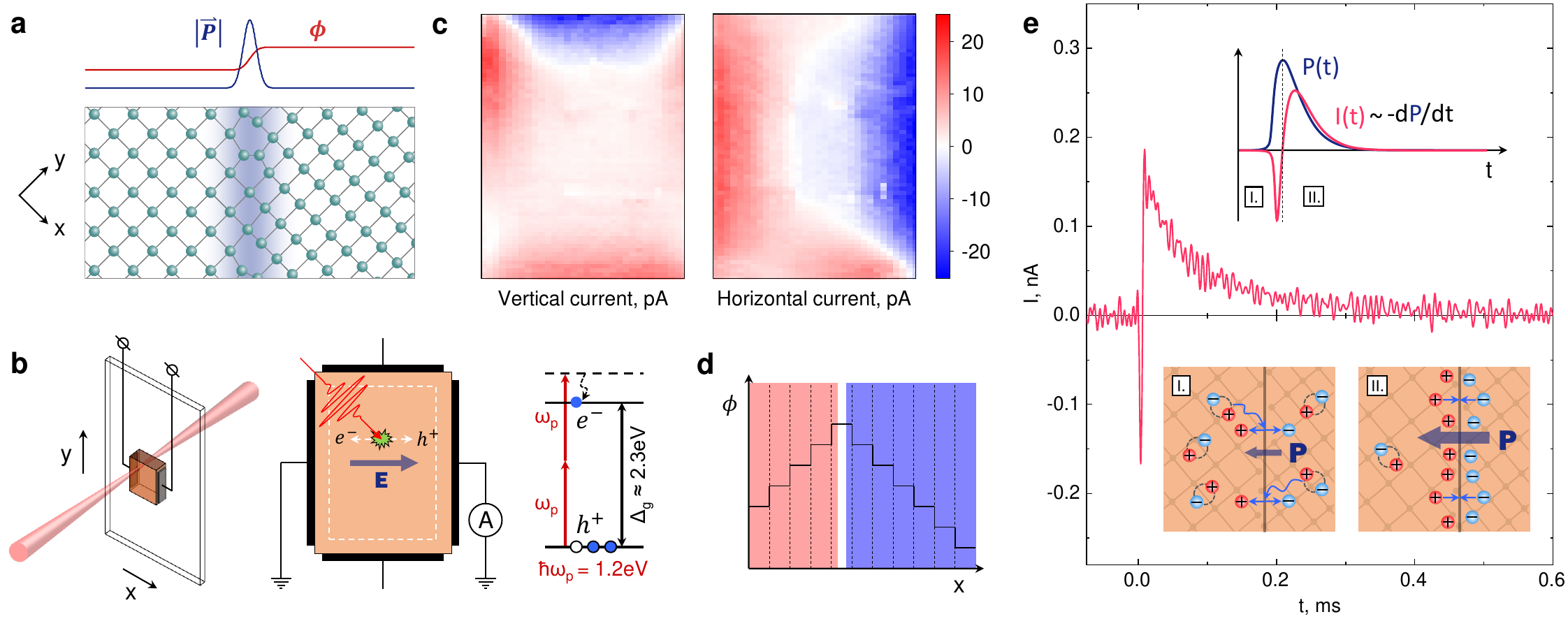}
\caption{
\textbf{Inversion symmetry breaking in MAPbBr$_3$ evidenced by the two-photon bulk photovoltaic effect.}
\textbf{a}, Schematic of the strain gradient confined to the domain wall, and the resulting gradients of polarization $P_{xy}$ and potential $\phi_{xy}$ induced by the flexoelectric effect.
\textbf{b}, Left: schematic of the photocurrent measurement setup. The sample is mounted on a glass slide and can be moved relative to the beam to measure the current flowing in both vertical and horizontal directions for each position of the excitation spot. A pair of contacts used to measure the photocurrent flowing in a horizontal direction is shown. Middle: schematic of the photoexcited carrier separation caused by internal electric fields. To measure the photocurrent in either the horizontal (as shown) or vertical direction, the corresponding pair of contacts was connected directly to the current input of the lock-in amplifier, leaving the other pair of contacts open. The dashed white line marks the part of the sample where the current measurements were performed. Right: energy diagram of the two-photon absorption process.
\textbf{c}, Spatial distribution of the photocurrent. Each point represents the current measured in either vertical or horizontal direction when the sample is excited at the corresponding location.
\textbf{d}, Schematic of electrostatic potential distribution $\phi(x)$ in a sample with domain-wall flexoelectricity. The dashed lines represent domain walls.
\textbf{e}, Typical time-resolved photocurrent transient acquired in a horizontal direction. The top inset illustrates the temporal evolution of polarization P and current I following optical excitation. The bottom inset shows schematic microscopic pictures of the charge separation and recombination phases of the photocurrent generation. See text for details.
}
\label{fig:fig3}
\end{figure*}

\subsection{Bulk photovoltaic effect}

Strain gradient breaks inversion symmetry and in general results in electric polarization in a phenomenon known as flexoelectricity \cite{Zubko2013}. In light of recent reports of large voltages generated in response to externally-induced mechanical deformations in MAPbBr$_3$ \cite{Shu2020, Wang2024}, one may wonder if gradients of spontaneous strain in nominally cubic LHPs can also give rise to local electric polarization (Fig.\ref{fig:fig3}a).

Here, we confirm the presence of local electric polarization in bulk MAPbBr$_3$ by detecting zero-bias photocurrent in single-crystal samples following localized photocarrier injection. To this end, an ultrafast sub-bandgap laser pulse is focused inside the sample, generating electron-hole pairs deep inside the bulk through two-photon absorption (Fig.\ref{fig:fig3}b). The resultant photocurrent is picked up through non-metallic carbon leads on the edges of the sample (Fig.\ref{fig:fig3}b, for vertical and horizontal currents) by a lock-in amplifier in current detection mode (see Methods). In Fig.\ref{fig:fig3}c we show the maps of the current frequency component corresponding to the laser pulse repetition rate ($f_{rep} = 1.5$kHz) $I_{rep}$ obtained by scanning the beam across the sample. As seen in Fig.\ref{fig:fig3}c, the current $I_{rep}$ strongly depends on the location of carrier injection. It is evident that the sign of $I_{rep}$ remains constant across large sections of the sample which are not symmetrical with respect to the midplane of the sample, indicating that the photocurrent is likely not due to dynamic flexoelectricity \cite{Nova2019}. Instead, it can be understood in terms of electrostatic potential distribution $\phi(x)$ in a sample with domain-wall flexoelectricity as shown schematically in Fig.\ref{fig:fig3}d. Here the plateaus represent regions of constant strain and $\phi(x)$ within domains that both change abruptly at domain walls; the sign of photocurrent $I_{rep}$ is then determined by the average slope of electrostatic potential.

This picture offers a natural explanation to the conflicting phenomenology of the apparent ferroelectricity in LHPs: on the one hand, flexoelectric polarization at domain walls related to the difference in structure between domains naturally explains the observed pyroelectric phenomena at structural phase transitions (\textit{e.g.} \cite{Gao2019}) and the pinning of domains walls can account for hysteretic polarization under external electric field \cite{Rakita2017, Garten2019}. On the other, flexoelectric polarization remains confined to the domain walls, keeping inversion symmetry intact in bulk, in full agreement with optical second harmonic generation experiments \cite{Hirori2019, Ahmadi2018}.

\begin{figure*}
\includegraphics[width=\textwidth]{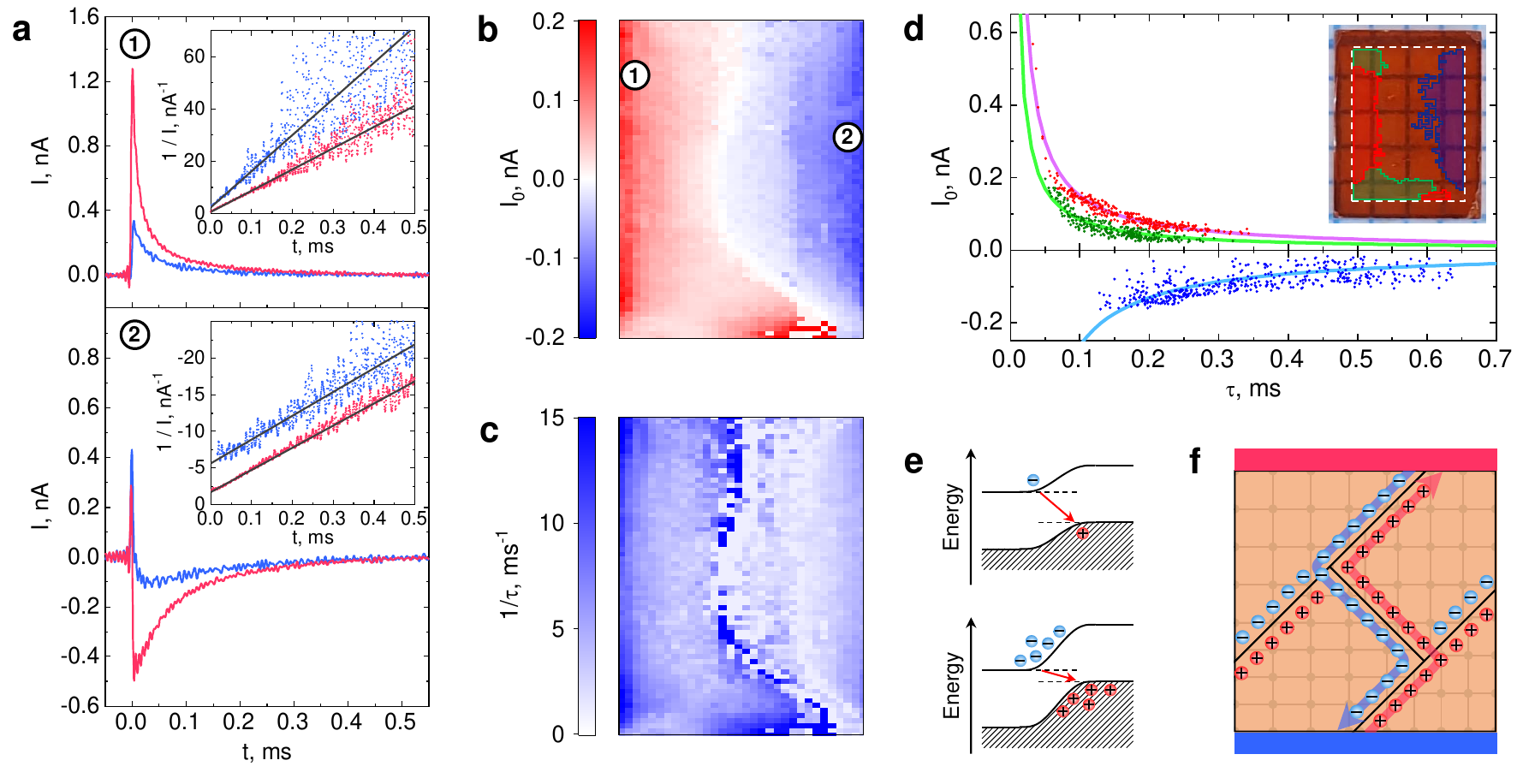}
\caption{
\textbf{Charge carrier dynamics in MAPbBr$_3$.}
\textbf{a}, Photocurrent transients measured in the horizontal direction at the respective locations indicated in \textbf{b}. Different colors correspond to measurements performed on a sample isolated from stray light (blue) and under diffuse ambient illumination of 0.01 sun (pink), respectively. Insets show the corresponding data fitted to $1/I(t) \propto (t + \mathrm{const})$ (black lines).
\textbf{b,c}, Spatial distribution of $I_0$ and $1/\tau$ obtained from exponential fitting of current decays in regime II at respective coordinates.
\textbf{d}, Correlation between $I_0$ and $\tau$. Three identified data clusters are fitted with $I_0 = C \cdot 1/\tau$, where $C$ is constant within each cluster. The inset shows a photograph of the sample, with highlighted regions corresponding to the identified clusters. The dashed white line marks the part of the sample where current measurements were taken.
\textbf{e}, Schematic of bandgap reduction caused by the accumulation of charge carriers at domain walls.
\textbf{f}, Schematic of charged domain walls acting as separate, charge-specific conductive pathways for electrons and holes in a perovskite solar cell.
}
\label{fig:fig4}
\end{figure*}

\section{Discussion}

Flexoelectric polarization at domain walls is responsible for photocurrent, but directly assigning it to simple charge diffusion along average slopes of potential in Fig.\ref{fig:fig3}d is problematic. Indeed, imagine electron-hole pairs injected \textit{e.g.} in the blue region in Fig.\ref{fig:fig3}d. According to the naive picture, the holes will diffuse to the right electrode, while electrons will accumulate at the peak position of $\phi(x)$ (marked with a white stripe), which cannot be sustained indefinitely. More generally, in the absence of external bias, the current in a charge-neutral insulating sample is given by $I(t) \propto -{dP}/{dt}$, where $P$ is electric polarization (Fig.\ref{fig:fig3}e). Therefore, if $\int I(t) dt \neq 0$ for each pulse, then the total electric polarization of the sample monotonically changes throughout the course of experiment, contradicting the observation that photocurrent in LHP samples can be generated for hours with no appreciable changes in magnitude.

To interpret our observations, we first note that, based on the preceding arguments, it must be that $\int I(t) dt = 0$ or, equivalently $P_{\mathrm{initial}} = P_{\mathrm{final}}$ (see upper inset in Fig.\ref{fig:fig3}e) which is in apparent contradiction with data in Fig.\ref{fig:fig3}e, showing transient current response $I(t)$ reconstructed as a function of time after the laser pulse (see Methods). We ascribe this to the bandwidth limitations of our lock-in measurements ($f \lesssim 153$kHz), because of which the high-frequency features like the fast spike near $t = 0$ in Fig.\ref{fig:fig3}e go under-sampled. Next, we note that although the majority of photocarriers recombine rapidly, a small fraction persists to produce the long-lived photocurrent, whose mechanism can be understood by scrutinizing the time dependence of a typical current transient $I(t)$ as in Fig.\ref{fig:fig3}e. One can identify here two distinct regimes: I) quick accumulation of polarization (a few $\mu$s) manifested as strong and narrow initial spike in $I(t)$, on the limit of time resolution of lock-in; and II) slow relaxation of polarization ($\sim 100\mu$s) manifested by the slow ``tail'' in $I(t)$ opposite in sign to the initial spike, providing the main contribution to the current picked up by lock-in. Microscopically, in regime I photocarriers diffuse to the domain boundaries where local electric fields spatially separate positive and negative charges, trapping them on the opposite sides of the domain wall; in regime II the polarization accumulated at the domain walls is gradually relaxing via tunneling through the flexoelectric potential barrier (Fig.\ref{fig:fig3}e).

The tunneling-driven recombination mechanism naturally reconciles the apparent paradox between the nanosecond-scale exciton lifetimes in perovskites \cite{Becker2018} and the millisecond-scale relaxation of photocurrent, a key factor in the photovoltaic performance of LHPs \cite{Huang2015}. Surprisingly, in regime II, the current $I(t)$ is best described not by a simple or even bi-exponential decay \cite{supp}, but rather by

\begin{equation}  
I(t) \propto \frac{1}{t + \mathrm{const}}  
\label{eq:current_law}  
\end{equation}  

\noindent as shown in the insets of Fig.\ref{fig:fig4}a. This time dependence is further supported by an unusual correlation between the current magnitude $I_0$ (Fig.\ref{fig:fig4}b) and its decay rate $1/\tau$ (Fig.\ref{fig:fig4}c) when fitting $I(t)$ in regime II with an exponential function $I(t) = I_0\cdot\exp\left( -t/\tau \right )$ \cite{supp}. Across extended regions of the sample, we find a strong correlation in the form $I_0\cdot\tau \approx \text{const}$ (Fig.\ref{fig:fig4}d). While such behavior is highly unexpected for a generic exponential decay, it emerges naturally from the form of $I(t)$ in eq.\ref{eq:current_law}.

Establishing the actual decay law of $I(t) \propto -\dot{P}$ reveals a more nuanced picture of relaxation dynamics of charge carriers $n(t) \propto P$ trapped at the domain wall. The behavior in eq.\ref{eq:current_law} suggests that $n(t)\propto \log(t+\mathrm{const})$. To understand this unexpected behavior, we notice that it is a solution of a modified kinetic equation, with density-dependent relaxation rate $\gamma(n)$:

\begin{equation}
{dn}/{dt} = -\gamma(n) n \equiv -\gamma_0 \exp\left( n/\bar{n} \right) n
\end{equation}

\noindent for $n(t)\gtrsim \bar{n}$. The exponential sensitivity of tunneling rate through the potential barrier on $n$ is natural since accumulation of electrons and holes on opposite sides of the domain walls leads to a decrease in the effective band gap and subsequent exponential growth of the tunneling rate (Fig. \ref{fig:fig4}e). General consideration yield $\bar{n} \sim 0.1/a^2$, $a$ standing for perovskite lattice constant (\cite{supp}; also \cite{Rappe2015}). The exponential sensitivity of tunneling on domain wall charging also explains the otherwise puzzling extreme sensitivity of relaxation rate and magnitude of photocurrent to ambient light shown in Fig.\ref{fig:fig4}a (red \textit{vs.} blue curves). Indeed, weak intrinsic recombination rates imply that even small amounts of stray light can significantly modify photocurrent transients.

The spatial separation of electrons and holes at domain walls strongly suppresses recombination, exponentially enhancing the effective lifetimes and thereby allowing charge carriers to diffuse over large distances. In this sense, charged domain walls can serve as effective conductive pathways for electrons and holes (Fig.\ref{fig:fig4}f), akin to the “ferroelectric highways” proposed by Frost \textit{et al.} \cite{Frost2014} in their comprehensive study on the potential ferroelectric ordering in LHPs.

\section{Conclusion}
In summary, we present evidence of localized inversion symmetry breaking in the nominally cubic high-temperature phase of MAPbBr$_3$, which is explicitly confirmed by measuring the two-photon bulk photovoltaic effect. We show that the generation of short-circuit photocurrent is driven by the flexoelectric effect resulting from strain gradients confined to domain boundaries. We visualize the emerging domain structure using an original technique based on electrochemical staining of the domain walls. Our results indicate that electric fields produced by localized flexoelectricity facilitate the accumulation of electrons and holes on opposite sides of the domain walls, which reduces recombination through the spatial segregation of charge carriers. The charged domain walls thus serve as efficient transport channels, extending the lifetime and diffusion lengths of the charge carriers. By revealing the peculiarities of carrier dynamics, we demonstrate that the unique properties of lead-halide perovskites are a natural consequence of their complex structure. Our findings unify seemingly contradictory observations into a coherent framework, providing a promising pathway for the design and optimization of hybrid perovskites in photovoltaic applications.

\section{Methods}

\textbf{Materials.} CH$_3$NH$_3$Br$_3$ (>99.99\%) was purchased from GreatCell Solar Ltd. (formerly Dyesol) and used as received. PbBr$_2$ (98\%), and DMF (anhydrous, 99.8\%) were purchased from Sigma Aldrich and used as received.

\textbf{Growth of CH$_3$NH$_3$PbBr perovskite single crystals.} A 1.5 M solution of CH$_3$NH$_3$Br/PbBr$_2$ in DMF was prepared, filtered through a 0.45 $\mu$m-pore-size PTFE filter, and the vial containing 0.5-1 ml of the solution was placed on a hot plate at 30$^\circ$C. Then the solution was gradually heated to 60$^\circ$C and kept at this temperature until the formation of CH$_3$NH$_3$PbBr$_3$ crystals. The crystals were collected and cleaned using a Kimwipe paper.

\textbf{Device fabrication.} For domain wall visualization, wires were attached to the single crystal samples using 8330D silver conductive epoxy (MG Chemicals). The epoxy was cured at room temperature for 24 hours. For photocurrent measurements, the single crystal sample was mounted on a glass holder \cite{supp}, and the wires were attached to it using flexible carbon conductive epoxy G6E-FRP (Graphene Laboratories, Inc.). After curing at room temperature for 24 hours, the epoxy adhesive provided both mechanical fixation and reliable electrical contact. The edges of the sample were covered with non-conductive epoxy to ensure isolation between the adjacent contacts.

\textbf{Birefringence measurements.} To visualize the natural birefringence in MAPbBr$_3$, the sample was put in a rotating crossed polarizer setup and illuminated with an expanded He-Ne laser beam (Thorlabs HNL050LB). The beam polarization was adjusted using a half-wave plate to match the orientation of the first polarizer (P1 in Fig.\ref{fig:fig1}c). The transmission images of the sample were acquired for each position of the crossed polarizer using a CMOS camera (Allied Vision Alvium 1800 U-319m).

To study the critical behavior of birefringence across the cubic-to-tetragonal phase transition, a near-infrared ($\lambda=1028$nm) beam polarized at $45\deg$ relative to [010] crystal axis was sent through the sample placed inside an optical cryostat normal to the (100) crystal face. At each temperature, a complete set of Stokes parameters was measured using the standard procedure \cite{Collett2005}. Special care was taken to compensate for thermal-expansion-driven displacement of the sample during the experiment \cite{supp}.

\textbf{Microscopy.} Bright-field and confocal microscopy were performed using a Zeiss LSM 880 Confocal Laser Scanning inverted microscope equipped with a Zeiss Plan-Apochromat 10x, 0.45 NA objective and PMT detectors for both confocal and transmitted light imaging. A He-Ne laser with $\lambda=632.8n$m was used for illumination. The sample was placed in a Petri dish with a glass coverslip bottom. The refractive index matching liquid Immersol 518F (Carl Zeiss Jena GmbH) was introduced between the sample and the coverslip to reduce reflection from the sample surface. Optical sectioning was performed with 5$\mu$m resolution along the optical axis and lateral resolution of 1.66$\mu$m per pixel. Illumination, focusing, optical sectioning, and primary image acquisition were controlled by Zeiss ZEN 2.3 SP1 Black software. Images were further processed using Zeiss ZEN 2.3 SP1 Black, NIH ImageJ 1.54p, and Helicon Focus 8.3.0 software.

\textbf{Photocurrent measurements.} Laser pulse source used was Light Conversion Pharos HP, 2$m$J per pulse at 3kHz repetition rate, central wavelength $\lambda=1028n$m and pulse duration $\tau=290f$. Only small part of the full laser power was used for the experiment to avoid sample damage. The actual average pump laser power was W=0.25mW at 1.5kHz pulse repetition rate after built-in laser pulse picker. Laser was focused with a lens with focal distance F=200mm, beam waist diameter inside sample being $w=40\mu$m.
To produce the spatially-resolved photocurrent maps the pump laser spot was scanned over the surface of the sample (see Fig.\ref{fig:fig3}b) with two pairs of contacts attached to the sample, enabling detection of the photocurrent flowing in vertical and horizontal directions. The measurements were performed using a lock-in amplifier (Zurich Instruments MFLI) in current mode (no bias), referenced to the laser output. The scanned area was limited to the part of the crystal away from the electrodes, eliminating the electrode proximity effects and excluding the possibility for the charge carrier to reach the contacts by diffusion. The pump polarization had no effect on the detected photocurrent.

Time-resolved transients of photocurrent were reconstructed from the current frequency components acquired by lock-in at integer multiples of the laser repetition rate (up to $n=102$ for high-resolution curves such as in Fig.\ref{fig:fig3}e and Fig.\ref{fig:fig4}a, and $n=18$ for area scans in Figs.\ref{fig:fig4}b,c). The amplitude and phase of each harmonic of the photocurrent were measured separately, and the time-domain representation of the signal was then reconstructed by inverse Fourier transform of the frequency domain data
\[
I(t) = \sum_{k=1}^{n} \left[ X_k \cos(2\pi f_k t) - Y_k \sin(2\pi f_k t) \right]
\]
where $X_k$ and $Y_k$ are real and imaginary parts of a k-th harmonic $f_k$ of the laser repetition rate, respectively. For the purpose of time-domain analysis of the current transients, the DC component appearing in the reconstructed signal is subtracted. Further information can be found in \cite{supp}.

\textbf{Data availability.} The data supporting the findings of this study are available from the corresponding author upon reasonable request.

\bibliography{main}
\end{document}